\documentclass[%
	floatfix,%
	reprint,%
	superscriptaddress,%
	nofootinbib,%
	twocolumn,%
	amsmath,%
	amssymb,%
	aps,%
	prl%
]{revtex4-2}

\usepackage{graphicx}
\usepackage{siunitx}
\usepackage{upgreek}

\begin{document} 

\title{Revealing Nonclassicality of Multiphoton Optical Beams \\ via Artificial Neural Networks}

\author{Radek Machulka}
	\email{radek.machulka@fzu.cz}
	\affiliation{
        Institute of Physics of the Academy of Sciences of the Czech Republic,
        Joint Laboratory of Optics of Palacký University and Institute of Physics AS CR,
        17. listopadu 50a, 772 07 Olomouc, Czech Republic
	}

\author{Jan Peřina, Jr}
	\email{jan.perina.jr@upol.cz}
	\affiliation{
		Joint Laboratory of Optics of
		Palacký University and
		Institute of Physics of AS CR,
		Faculty of Science, Palacký University,
		17. listopadu 12, 77146 Olomouc, Czech Republic
	}

\author{Václav Michálek}
	\affiliation{
        Institute of Physics of the Academy of Sciences of the Czech Republic,
        Joint Laboratory of Optics of Palacký University and Institute of Physics AS CR,
        17. listopadu 50a, 772 07 Olomouc, Czech Republic
	}

\author{Roberto de J. León-Montiel}
	\email{roberto.leon@nucleares.unam.mx}
	\affiliation{
		Instituto de Ciencias Nucleares,
		Universidad Nacional Autónoma de México,
		Apartado Postal 70-543, 04510 Cd. Mx., México
	}

\author{Ondřej Haderka}
	\affiliation{
		Joint Laboratory of Optics of
		Palacký University and
		Institute of Physics of AS CR,
		Faculty of Science, Palacký University,
		17. listopadu 12, 77146 Olomouc, Czech Republic
	}

\begin{abstract}
The identification of nonclassical features of multiphoton quantum states represents a task of the utmost importance in the development of many quantum photonic technologies. Under realistic experimental conditions, a photonic quantum state gets affected by its interaction with several non-ideal opto-electronic devices, including those used to guide, detect or characterize it. The result of such noisy interaction is that the nonclassical features of the original quantum state get considerably reduced or are completely absent in the detected, final state. In this work, the self-learning features of artificial neural networks are exploited to experimentally show that the nonclassicality of multiphoton quantum states can be assessed and fully characterized, even in the cases in which the nonclassical features are concealed by the measuring devices. Our work paves the way toward artificial-intelligence-assisted experimental-setup characterization, as well as \emph{smart} quantum-state nonclassicality identification.
\end{abstract}

\maketitle

\section{Introduction} 

\begin{figure*}[t!]
\centering
	\includegraphics[width=0.9\textwidth]{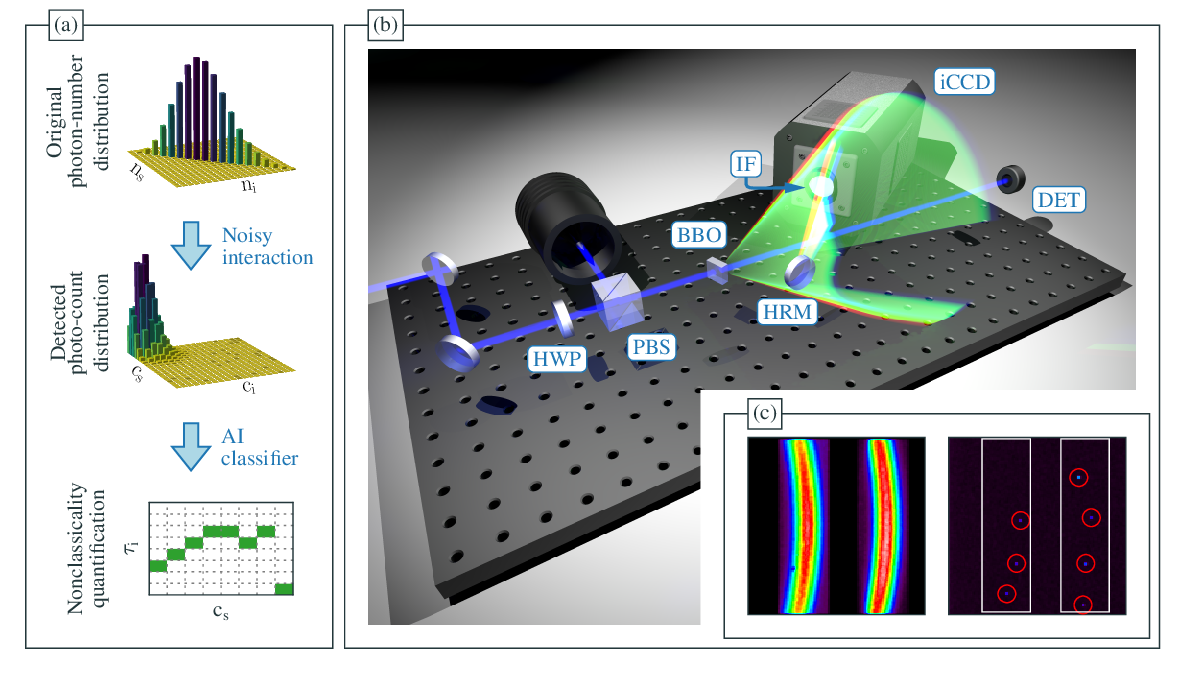}
	\caption{Smart nonclassicality quantification. (a)~Flow diagram of the proposed artificial-intelligence-based method for classification and quantification of nonclassicality of multiphoton optical beams. (b)~Schematic representation of our experimental setup. Twin beams, distributed in a cone-shaped region, are generated by means of spontaneous parametric down-conversion (SPDC) in a second-order, nonlinear crystal pumped by a strong laser beam. A small part of the signal is detected with a camera directly, while the conjugated section of the cone is reflected on the camera by a high-reflecting mirror. The labels correspond to: HWP~-~half-wave-plate, PBS~-~polarizing beam-splitter, BBO~-~$\upbeta$-BaB$_{2}$O$_{4}$ nonlinear crystal, HRM~-~high-reflectivity mirror, DET~-~detector, IF~-~interference filter, iCCD~-~intensified CCD camera. (c) Detector images: left panel~-~single shot image with both signal and idler regions of interest (ROIs) (white rectangles) as well as individual detection events (red circles). Right panel~-~accumulated detection events identified in~\num{1.2e6} individual single shot frames.}
\label{Figure1}
\end{figure*}

Nonclassical states of light were unveiled by Glauber and Sudarshan in 1963, through the so-called Glauber-Sudarshan state representation~\cite{Glauber1963, Sudarshan1963}. This fundamental tool allows one to uniquely distinguish between the classical and quantum states of light~\cite{Mandel1995}. Further studies of nonclassical states revealed their unique properties in two broad areas: Reduced fluctuations of measured quantities below their sub-shot-noise level~\cite{Dodonov2002, Luks1988}, and the presence of various kinds of quantum correlations~\cite{Hill1997, Adesso2007} among subsystems (entanglement). While the former property finds its applications in quantum (ultra-precise) metrology, namely the use of squeezed light in the measurement of relative phase~\cite{Lvovsky2009}, sub-Poissonian light in absorption measurements~\cite{Li2020}, and quantum imaging~\cite{Genovese2016}; the latter is substantial for quantum communication and quantum information protocols~\cite{Zeshen2023}, as well as quantum measurements~\cite{Klyshko1980}. Importantly, these particular properties are manifestations of one common feature called \emph{nonclassicality}, which is a very general feature of quantum systems and, as such, it is very difficult to quantify.

Following textbook definitions, nonclassicality is, arguably, a synonym for negative values of the quasi-distribution of field amplitudes in phase space~\cite{Glauber1963, Sudarshan1963, Perina1990}, or even its non existence as a regular function. Roughly speaking, the more negative the quasi-distribution becomes, the more nonclassical the field is. One of the nonclassicality quantifiers derived directly from its definition receives the name of nonclassicality volume~(NV). Introduced by Kenfack and Zyczkowski~\cite{Kenfack2004}, NV evaluates the volume between the negative values of the Wigner quasi-distribution function related to the symmetric field-operator ordering and zero. Unfortunately, this metric may only be applied to quantum states that express their nonclassicality around the symmetric field-operator ordering. More importantly, it requires the reconstruction of the Wigner function through experimentally demanding homodyne tomography~\cite{Lvovsky2009}.

Because of the above, alternative approaches for nonclassicality identification have been developed using various kinds of nonclassicality inequalities~\cite{Richter2002, Miranowicz2010, Ryl2015}. These inequalities emerge from negation of classical inequalities, such as the Cauchy-Schwarz inequality~\cite{agarwal1988}, or by considering various kinds of nonnegative quadratic forms of polynomials or relying on the majorization theory (for details, see Refs.~\cite{PerinaJr2017a, PerinaJr2020a, PerinaJr2022}). Such inequalities are typically referred to as nonclassicality witnesses~(NCWs), and are written in terms of the field intensity moments~\cite{Perina1991}. Interestingly, by using the Mandel detection formula~\cite{Perina1991}, NCWs can be transformed into inequalities written in terms of probabilities of photon-number distributions~\cite{Klyshko1996,PerinaJr2022}. Notably, by using the concept of Lee nonclassicality depth~(NCD)~\cite{Lee1991}, a corresponding NCD may be assigned to each NCW. This leads to a systematic form of quantification of nonclassicality.

When studying nonclassicality, one immediately runs into the problem that there does not exist a~universal NCW capable of identifying all nonclassical states in a~given quantum system. For each kind of quantum state there exist specific NCWs suitable for identification and quantification of its nonclassicality~\cite{PerinaJr2017a, PerinaJr2020a, PerinaJr2019}. In principle, all known NCWs may be applied and the corresponding NCDs determined. Suitable NCWs are then recognized by the greatest NCDs. Such approach has been applied to quantum 1D~\cite{PerinaJr2019}, 2D~\cite{PerinaJr2017a, PerinaJr2020a} as well as 3D fields~\cite{PerinaJr2021, PerinaJr2021a}. In these examples, multiphoton quantum fields are first characterized by photon-number-resolving measurements to obtain the experimental photo-count histograms. Note that these histograms are obtained after the original quantum state interacts with all optical elements (including detector) placed before the actual measurements. Then, by using different kinds of reconstruction methods, the true photon-number distributions of the measured fields, as well as the quantities needed for the NCWs, are extracted. It is worth pointing out that this standard approach is computationally expensive as it requires complex reconstruction algorithms, as well as evaluation of a~great number of NCWs to determine their corresponding NCDs.

Notably, artificial intelligence algorithms have progressively revolutionized a wide range of disciplines, and quantum optics is no exception. Indeed, artificial neural networks have successfully been used in the classification of quantum states~\cite{ahmed2021} and light sources~\cite{you1}; in the verification of nonclassicality of states probed via homodyne detection~\cite{gebhart2021}, in super-resolved quantum imaging~\cite{Bhusal2022}, in the quantification of the degradation of squeezed states in balanced homodyne detection~\cite{hsieh2022}, and in the measurement of entanglement in many-body systems~\cite{gray2018,Asif2023,koutny2023}. In this work, we demonstrate that artificial neural networks can effectively be used to assess and quantify nonclassicality of multiphoton quantum states, even in scenarios where nonclassical features are affected, or completely destroyed, by the measuring devices. Our results thus pave the way toward artificial-intelligence-assisted experimental-setup characterization, as well as smart quantum-state nonclassicality identification.

\section{Results} 

\noindent \textbf{Theoretical Background}

\noindent Our nonclassicality-quantification machine-learning approach, schematically shown in Fig.~\ref{Figure1}(a), considers a group of one-dimensional quantum fields with sub-Poissonian photon-number distributions~\cite{PerinaJr2013b, Arkhipov2016c}. Such states arise in the process of post-selection, which is managed by means of photon-number-resolving detection of one of the modes of the so-called twin-beam (TWB) state~\cite{PerinaJr2017c}. Twin beams are experimentally generated by the process of spontaneous parametric down-conversion, as illustrated in Fig.~\ref{Figure1}(b). Note that in the ideal photon-number-resolved detection of an~ideal TWB, if the detection on one beam gives~$n_{s}$ photo-counts, the remaining beam is left in the Fock state with~$n_{i}=n_{s}$ photons, i.e. in a~highly nonclassical state. It is worth remarking, however, that in a realistic photon-number-resolving detector, with a finite detection efficiency, photon-number distribution of the ideal Fock state is affected, but it may still remain sub-Poissonian~\cite{Laurat2003,Brida2012,PerinaJr2013b}, i.e. with a nonzero NCD.

The standard method for estimating nonclassicality of multiphoton beams via NCD proceeds as follows. In the experiment, a photo-count histogram~$f_{\rm i}(c_{\rm i}; c_{\rm s})$ characterizing an~idler field obtained after detection of~$c_{\rm s}$ photo-counts in the signal field is recorded. Note that indices~s and~i stand for the TWB's signal and idler fields, respectively. The histogram is connected to the idler-field photon-number distribution~$p_{\rm i}(n_{\rm i}; c_{\rm s})$ through the following expression~\cite{Haderka2005, PerinaJr2012}:
\begin{equation}
	f_{\rm i}(c_{\rm i}; c_{\rm s}) =
	\sum_{n_{\rm i} = 0}^{\infty}
	T_{\rm i}(c_{\rm i}, n_{\rm i}) p_{\rm i}(n_{\rm i}; c_{\rm s}),
\label{eq1}
\end{equation}
where the matrix~$T_{\rm i}(c_{\rm i}, n_{\rm i})$ describes the realistic detection process used for monitoring the idler field. In other words, it provides the probability of detecting~$c_{\rm i}$ photo-counts after $n_{\rm i}$ photons have arrived to the detector.

The detection matrix~$T_{\rm i}$ is assumed to be known and the relation in equation~(\ref{eq1}) is inverted in the form of the following iteration formula using the maximum-likelihood estimation~(MLE) approach~\cite{Dempster1977,Haderka2005} ($j=1,2,\ldots$):
\begin{eqnarray}
	p^{(j+1)}_{\rm i}(n_{\rm i};c_{\rm s}) &=&
		p^{(j)}_{\rm i}(n_{\rm i};c_{\rm s}) \nonumber \\
		&\times& \sum_{c_{\rm i}=0}^{\infty}
		\frac{f_{\rm i}(c_{\rm i};c_{\rm s}) T_{\rm i}(c_{\rm i},n_{\rm i})}
		{\sum_{n'_{\rm i}=0}^{\infty} T_{\rm i}(c_{\rm i},n'_{\rm i})
			p^{(j)}_{\rm i}(n'_{\rm i};c_{\rm s})}.
\label{eq2}
\end{eqnarray}
The photon-number distribution~$p_{\rm i}(n_{\rm i};c_{\rm s})$ is revealed as a~steady state of the iteration algorithm represented by equation~(\ref{eq2}). It is then analyzed from the point of view of nonclassicality.

For one-dimensional fields, suitable NCWs are given by~\cite{Arkhipov2016c, PerinaJr2017a}
\begin{equation}
	L^{kl}_{mn} =
	\langle W_{\rm i}^k \rangle
	\langle W_{\rm i}^l \rangle -
	\langle W_{\rm i}^m \rangle
	\langle W_{\rm i}^n \rangle < 0,
\label{eq3}
\end{equation}
where $\langle W_{\rm i}^k \rangle$ stands for idler-field intensity moments of order $k$. Note that intensity moments are normally-ordered photon-number moments (see Methods Section for details). As we discuss below, the order up to which equation~(\ref{eq3}) is evaluated, depends directly on the experimental conditions and errors. Note that by making use of Mandel's detection formula~\cite{Perina1991,Klyshko1996,PerinaJr2017a}, the intensity NCWs can be transformed into probability NCWs (see Methods Section for details) defined as~\cite{PerinaJr2017c, PerinaJr2020}:
\begin{equation}
	\bar{L}^{kl}_{mn} =
	k_{\rm i}! l_{\rm i}! p_{\rm i}(k_{\rm i}) p_{\rm i}(l_{\rm i}) -
	m_{\rm i}! n_{\rm i}! p_{\rm i}(m_{\rm i}) p_{\rm i}(n_{\rm i}).
\label{eq4}
\end{equation}
It is worth pointing out that both, the intensity and probability NCWs play an important role in nonclassicality quantification, as they identify the possible nonclassical features in complementary domains, namely field-moments and photon-number distributions.

The strongest manifestation of the nonclassicality is observed in the natural, normal ordering of the field operators~\cite{Perina1991}. When moving from the normal ordering towards the anti-normal ordering of field operators, we effectively add an~additional 'field-operator ordering' detection noise that physically originates in spontaneous emission of electrons in the detector. The intensity of such signal increases linearly with the operator-ordering parameter, $s$, that quantifies the amount of spontaneous emission~\cite{Perina1991}. The increasing detection noise as we move outside the normal ordering gradually conceals the nonclassicality features~\cite{Lee1991} observed in the quasi-distributions of field amplitudes in the phase-space~\cite{Mandel1995}. At certain threshold value~$s_{\rm th}$, the nonclassicality features are completely concealed and we have a~classical distribution of field amplitudes. According to Lee~\cite{Lee1991}, this threshold value $s_{\rm th}$ allows to determine a quantifier of the nonclassicality that is called the Lee nonclassicality depth~(NCD)~\cite{Lee1991}
\begin{equation}
	\tau = (1-s_{\rm th})/2 .
\label{eq5}
\end{equation}
It attains the values in the interval~$\langle 0,1\rangle$.

Once we know how to transform the intensity moments and photon-number distributions from their usual form, i.e.~normal ordering, to an~arbitrary field-operator ordering~$s$~\cite{Perina1991} (see Methods Section for details), we may assign to any NCW the corresponding NCD. This is possible because the NCWs gradually lose their ability to identify the field nonclassicality as the field-operator ordering departs from the normal ordering. Each NCW loses completely its ability of nonclassicality identification at certain value of the ordering parameter~$s_{\rm th}$, i.e.~specific NCD~$\tau$ is assigned to any NCW along equation~(\ref{eq5}).

At this point, we have seen that standard evaluation of NCD~$\tau$ for an~arbitrary multiphoton field comprises two demanding tasks: reconstruction of photon-number distribution~\cite{Haderka2005, PerinaJr2012} and analysis of NCWs in either the intensity-moment space~\cite{Arkhipov2016c, PerinaJr2017a} or in photon-number distributions~\cite{PerinaJr2017c, PerinaJr2020}. Interestingly, machine learning algorithms can be used to combine the two steps into one by means of artificial neural networks~(ANNs). As one might expect, the problem of sufficient amount of training data for the ANN has to be solved. Indeed, due to practical limitations on the number of measurement realizations and repetitions, it is not possible to acquire sufficient number of experimental data. Nevertheless, as we show next, a~suitable theoretical model can be used to generate the training data instead.

\begin{figure*}[t!]
\centering
	\includegraphics[width=0.9\textwidth]{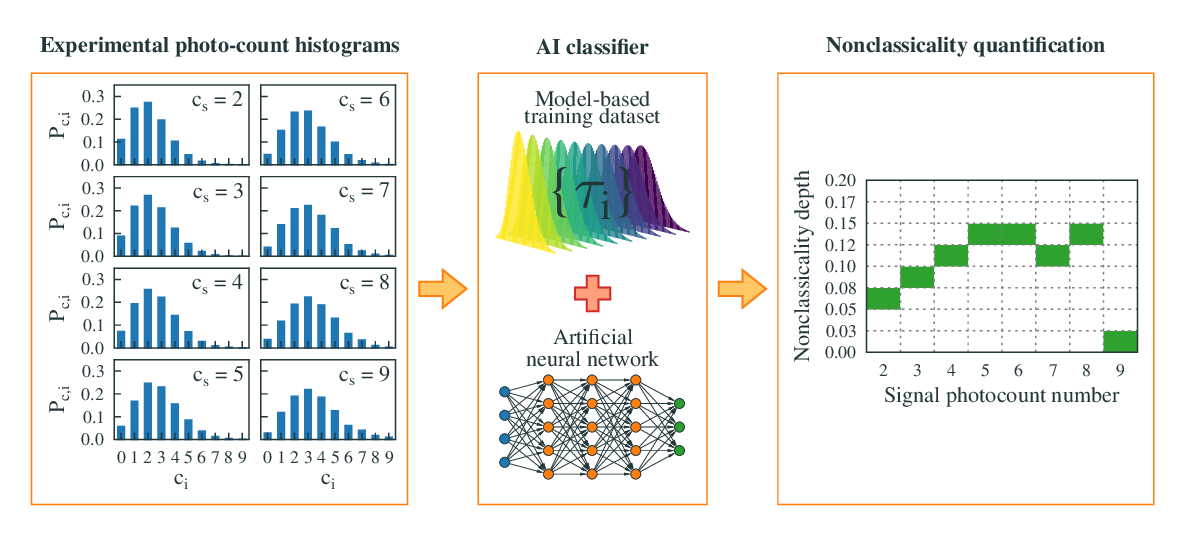}
	\caption{Artificial neural network working principle. Our neural networks are designed and trained using a set of model-based photo-count distributions, together with their nonclassicality depths~($\tau_{i}$). The trained network is then used to classify experimentally-obtained photo-count histograms.}
\label{Figure2}
\end{figure*}

The twin beams---which can be reconstructed from experimental histograms, $f_{\rm i}(c_{\rm i}; c_{\rm s})$---may be faithfully described by three independent multi-mode Gaussian fields corresponding to: ideal photon-pair component of the field, noise component of the signal field and noise component of the idler field. The photon-number distribution of such state, $p^{\rm TWB}(n_{\rm s}, n_{\rm i})$, can thus be written as~\cite{PerinaJr2013a}
\begin{eqnarray}
	p^{\rm TWB}(n_{\rm s}, n_{\rm i}) &=&
		\sum_{n = 0}^{\text{min}(n_{\rm s}, n_{\rm i})}
		p(n_{\rm s} - n; M_{\rm s}, B_{\rm s}) \nonumber \\
	&\times&
		p(n_{\rm i} - n; M_{\rm i}, B_{\rm i}) p(n; M_{\rm p}, B_{\rm p}),
\label{eq6}
\end{eqnarray}
in which the Mandel-Rice distribution~$p(n; M, B) = \Gamma(n + M)/[n! \Gamma(M)] B^{n} / ((1 + B)^{n + M}$ gives the photon-number distribution of an~$M$-mode field with~$B$ mean photon(-pairs) per mode. Here, the indices~p, s, and~i correspond to the photon-pair, signal-noise, and idler-noise field components, respectively. The theoretically-predicted photon-number distributions~$p^{\rm th}_{\rm i}(n_{\rm i}; c_{\rm s})$ of the idler fields are then written as
\begin{equation}
	p^{\rm th}_{\rm i}(n_{\rm i}; c_{\rm s}) =
	\sum_{n_{\rm s}=0}^{\infty}
	T_{\rm s}(c_{\rm s}, n_{\rm s}) p^{\rm TWB}(n_{\rm s}, n_{\rm i}),
\label{eq7}
\end{equation}
where the detection matrix~$T_{\rm s}(c_{\rm s}, n_{\rm s})$ describes the detector in the signal field. Note that the prediction~$f^{\rm th}_{\rm i}$ for the experimental histogram~$f_{\rm i}$ using the photon-number distribution~$p^{\rm th}_{\rm i}$ satisfies the relation given in equation~(\ref{eq1}).

The group of twin beams considered for training is characterized by six parameters~\cite{PerinaJr2013a}. However, two parameters are obtained in the experiment with very high precision, namely the mean photon numbers in the signal and idler fields, which are given by~$M_{\rm s} B_{\rm s} + M_{\rm p} B_{\rm p}$ and~$ M_{\rm i} B_{\rm i} + M_{\rm p} B_{\rm p} $, respectively. Moreover, it is well-known that the number~$M_{\rm p}$ of photon-pair modes of typical twin beams is usually in greater tens~\cite{PerinaJr2012}, and so their change influences the photon-number distributions negligibly. This analysis thus suggests that one only needs to vary the mean photon-pair number~$B_{\rm p}$ when training the ANN, and determine the mean noise signal and idler photon numbers~$M_{\rm s} B_{\rm s}$ and~$M_{\rm i} B_{\rm i}$ from the above mentioned experimental parameters. Once the noise mode numbers~$M_{\rm s}$ and~$M_{\rm i}$ are substituted by experimentally, best-fitted values, the mean photon-pair number~$B_{\rm p}$ remains the only varying parameter to generate data with varying NCD for the ANN training.
\\
\\

\noindent \textbf{Experimental-data preparation}

\noindent The experimental setup used to generate multiphoton sub-Poissonian states in the idler mode is depicted in Fig.~\ref{Figure1}(b). The third harmonic generation~(THG) of a cavity-dumped Ti:Sapphire laser system is used to pump the nonlinear SPDC process in a~\qty{5}{mm}~long BBO crystal, cut at~\ang{48}, for Type~I non-collinear interaction. The laser system is tuned to the central wavelength of~\qty{840}{nm} (THG:~\qty{280}{nm}), producing nearly transform-limited, \qty{150}{fs}~long optical pulses with~\qty{0.1}{\uJ} energy (THG:~\qty{20}{nJ}) at repetition rate of~\qty{50}{kHz}. Because the data acquisition takes usually tens of hours, the intensity of the pump beam and therefore the number of generated photon pairs per pulse is actively stabilized using an~optical system comprising a half-wave-plate in a~motorized rotary stage, a linear polarizer and feedback loop from a~suitable photo-detector (for details, see Ref.~\cite{PerinaJr2017c}). The power fluctuations are limited well below~\qty{1}{\percent~RMS}. To allow the stabilizer to compensate for possible drops in the pump power, the stable point is chosen at the half of the nominal power.

The generated photon pairs are emitted in a~cone-like geometry with the external vertex half angle~\ang{13} constrained by the phase-matching condition. The signal part of the cone is detected directly by the camera, while the corresponding idler part of the twin beam is reflected to the camera by a highly-reflecting mirror. Both detected parts form a strip-like pattern in the respective regions of interest~(ROI) on the detector. For typical single shot image, as well as accumulated signal frame revealing the spatial distribution of a twin beam in the transverse plane, see Fig.~\ref{Figure1}(c). Prior detection, both beams are spectrally filtered using an~interference filter centered at \qty{560}{nm}, with a bandwidth of~\qty{14}{nm} at FWHM. This allows us to minimize the noise coming from fluorescence, as well as any stray light from the laboratory. Note that the spectral width of the filter also defines the fringe widths at the camera.

An intensified CCD camera (iCCD, Andor DH334-18U-63) in high-gain, low-noise mode is used for the detection. In general, iCCD cameras lack photon-number resolution in individual pixels and can serve only as massive, multichannel on-off detectors. The photon-number resolving power therefore comes from the spatial spread of an~incident field~\cite{PerinaJr2012}. The original resolution of 1024x1024 pixels is reduced via hardware binning to only 128x128 macro-pixels to increase the signal-to-noise ratio, as well as to reduce the amount of transferred data. This step increases the speed of data acquisition up to~\qty{10}{Hz}. Moreover, it reduces spatial blurring of single photon detection events caused by imperfect coupling between an~intensifier and a CCD chip in the camera. Finally, by using trigger pulses from the laser system, the camera is gated to detect events within a \qty{4}{ns} long time-window, which further limits the noise.

The detection matrix, $T$, of an~iCCD camera with detection efficiency~$\eta$, $N$~detecting macro-pixels and mean dark-count rate per pixel~$D$ can be expressed in the form~\cite{PerinaJr2012}
\begin{eqnarray}
	T(c, n) &=& \binom{N}{c} (1 - D)^N (1 - \eta)^n (-1)^{c} \nonumber \\
	&\times&
		\sum_{l = 0}^{c} \binom{c}{l} \frac{(-1)^{l}}{(1-D)^l}
		\left( 1 + \frac{l}{N} \frac{\eta}{1-\eta} \right) ^n.
\label{eq8}
\end{eqnarray}
To fully characterize our setup, we make use of equation~(\ref{eq8}), together with the method outlined in Ref.~\cite{PerinaJr2012a}, to find that the experimental histogram of our SPDC source, $h^{\rm TWB}(c_{\rm s}, c_{\rm i}) \equiv f(c_{\rm i}, c_{\rm s})$, best-fits a twin beam with the following parameters: $M_{\rm p} = 270$, $M_{\rm s} = 0.01$, $M_{\rm i} = 0.026$, $B_{\rm p} = 0.032$, $B_{\rm s} = 7.6$, and $B_{\rm i} = 5.3$ (with relative errors~\qty{7}{\percent}). The parameters characterizing our detector are thus estimated to be: $\eta_{\rm s} = 0.228 \pm 0.005$, $\eta_{\rm i} = 0.223 \pm 0.005$, $D_{\rm s} = 0.206$, $D_{\rm i} = 0.214$ and $N_{\rm s} = 6528$, $N_{\rm i} = 6784$.
\\
\\

\noindent \textbf{Artificial neural network~(ANN) architecture}

\noindent The determination of NCDs for experimental one-dimensional histograms relies on classification by means of supervised ANNs. In such scheme, depicted in Fig.~\ref{Figure2}, a~data set comprising photo-count histograms, and the corresponding NCDs, is generated using the theoretical model described above. More specifically, the training data are prepared by varying the mean photon-pair number~$B_{\rm p}$ per mode in the Gaussian multi-mode model [see equation~(\ref{eq7})], while keeping the number of modes~$M_{\rm p}$, $M_{\rm s}$ and~$M_{\rm i}$ fixed at appropriate values (see above). In addition, we use the mean signal- and idler-beam photon-numbers obtained from the experiment. Such data set is then applied to train a~suitable ANN, which in turn assigns appropriate NCDs to given experimental photo-count histograms.

Our classification approach is based on a fully connected, feed-forward, multi-layer perception ANN. The structure, illustrated in Fig.~\ref{Figure2}, is arranged as follows: The input layer consists of~40~neurons, where each neuron represents an~individual element (i.e. probability) in the analyzed photo-count histogram. It is worth pointing out that the number of elements is chosen to correctly allocate the information of fields with different intensities. Then, a~set of identical hidden layers with a~rectified linear unit activation function~(ReLU) are added. The best performance of the network is found to be with two hidden layers, each with~20~neurons, for intensity-moment based NCWs, whereas the network trained with probability based NCWs needed three hidden layers with~50~neurons each. Finally, an~output layer with softmax activation function is added to the network. We remark that the number of neurons in the output layer defines the resolution of the NCD classification method, and needs to be chosen to match the experimental errors, see next subsection for details.

To quantify the model progress during training, we make use of a~logarithmic loss function, namely the categorical cross entropy. This function is defined as~$L_{CE} = -\sum_{i=1}^{n}p_{i}\log(q_{i})$, where~$n$ represents the number of classes, $p_{i}$ is the expected result (i.e., the ground truth), and~$q_{i}$ is the output probability given by the softmax function. To find the minimum of the loss function, individual weights are adjusted following the adaptive moment estimation algorithm~(ADAM). It combines the stochastic gradient descend method (root mean squared propagation combined with adaptive gradients) with the (first and second order) momenta into the fast converging, computationally efficient optimization algorithm.
\begin{figure}[t!]
\centering
	\includegraphics{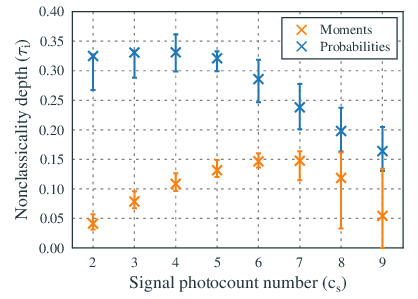}
	\caption{Nonclassicality depths from experimentally acquired photo-count histograms calculated using intensity moments and probability based NCWs from reconstruction using Gaussian twin-beam fit.}
\label{Figure3}
\end{figure}
\\
\\

\noindent \textbf{Optimized nonclassicality quantifiers}

\noindent During the optimization process of ANNs for NCD determination via intensity moments and probabilities of photo-count distributions, two different networks (one for each method) were found to be the best. The reason for this is that the NCWs based on intensity moments and probabilities, together with their corresponding NCDs, are sensitive to different, in certain sense complementary, features of the analyzed photo-count distributions that reflect their nonclassicality.

\begin{figure}[t!]
\centering
	\includegraphics{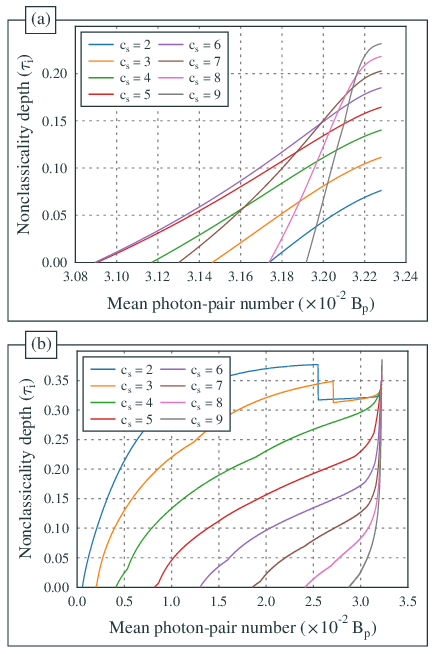}
	\caption{Nonclassical depths generated by training data for post-selected idler beams using~(a) intensity-moments NCWs and~(b) probability NCWs. The discontinuities in~(b) are caused by restriction on NCW set.}
\label{Figure4}
\end{figure}

\begin{figure}[t!]
\centering
	\includegraphics{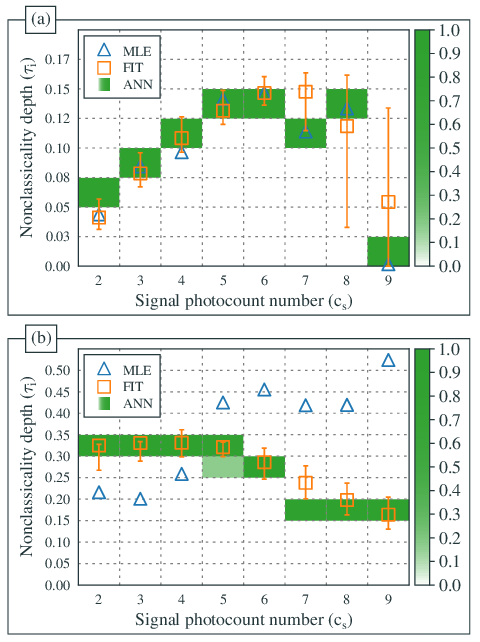}
	\caption{Nonclassicality depths, $\tau$, for post-selected idler beams determined from~(a) intensity NCWs and~(b) probability NCWs as predicted by artificial neural network~(ANN) are compared with those originating in MLE reconstruction method~(blue~$\triangle$) and obtained via the Gaussian FIT model~(orange~$\Box$). The green colorbar represents the classification success probability of a given input state.}
\label{Figure5}
\end{figure}

We first focus our attention on the determination of NCDs from intensity-moment NCWs. Taking into account that experimental errors increase with the increasing order of intensity moments, we consider the intensity-moment NCWs containing up to the third-order intensity moments. We note here that also the ability to reveal the state's nonclassicality and to assign an~appropriate NCD often decreases with the higher-order moments found in an~NCW. Even though the range of NCD values attained by NCWs with the intensity moments up to the third order is broader, the selected NCD values were limited to the interval~$\langle 0, 0.2 \rangle$, which safely covers the fields emerging in the experimental setup, see Fig.~\ref{Figure3}. This interval of expected NCDs was then divided into eight equidistant classes in accordance with the experimental precision. NCDs,~$\tau$, obtained during the preparation of the training data are plotted in Fig.~\ref{Figure4}(a). Because the NCD values are distributed unequally into the defined classes, the obtained training data is balanced to minimize potential bias of the network towards a particular NCD class. Once balanced, each class is covered by~\num{2e6} samples. Finally, after randomization, the data is used for training. From the whole dataset, we allocate~\qty{80}{\percent} for training and~\qty{20}{\percent} for validation. The trained ANN is then used to assign the corresponding NCDs~$\tau$ arising from intensity-moment NCWs to the set of eight~idler experimental histograms~$h_{\rm i}(c_{\rm i}; c_{\rm s})$ for~$c_{\rm s} \in \langle 2, 9\rangle$. As shown in Fig.~\ref{Figure5}(a), the values of NCDs~$\tau $ predicted by the trained ANN are in good agreement with the results found by applying both, the MLE method~[Eqs.~(\ref{eq1})-(\ref{eq2})], and the Gaussian multi-mode, twin-beam fitting~(FIT)~[Eqs.~(\ref{eq6})-(\ref{eq7})]. Notably, the ANN is capable of reconstructing specific local features of individual photo-count distributions. This is not the case of the Gaussian twin-beam fit to whom these specific features are invisible~\cite{PerinaJr2013a}. We would like to point out that the larger experimental uncertainties in NCDs with higher signal photo-count-numbers are caused by the small number of events that occur in these low-probability cases.

Similarly as for intensity-moment NCWs, a suitable ANN for the probability NCWs was designed and trained using the photo-count distributions and corresponding NCDs produced by theory. NCDs covered by training data are shown in Fig.~\ref{Figure4}(b). Note that the values of training-NCDs as well as experimental NCDs (shown in Fig.~\ref{Figure3}) are considerably greater than the intensity-moment case, and so the interval for expected NCDs was broadened to~$\langle 0, 0.5 \rangle$. Also, the experimental errors are distributed in a~different way as they increase with the decreasing probability. Given the obtained values for probabilities, and our measurement precision, the classification of different outcomes was divided in ten~intervals. The structure of the ANN is equivalent to the one for intensity-moment NCWs, and the same is true for the training process. The performance of the prepared ANN can be judged according to the results presented in Fig.~\ref{Figure5}(b). Note that the ANN results nearly coincide with those based on the FIT method. However, they systematically assign lower values of NCDs for weaker post-selected idler fields and greater values of NCDs for stronger idler fields compared to MLE. Lower values of NCDs provided by the MLE reconstruction method are a consequence of measurement-noise observed at low photo-count numbers, i.e., in the area where the nonclasicality of sub-Poissonian fields manifests itself [see~Figs.~3(a) and~5(a) in Ref.~\cite{PerinaJr2020}]. With the increasing idler field intensity, the area with the manifested nonclassicality moves towards greater photo-count numbers, which allows for its easier identification. On the other hand, larger values of NCDs from the MLE method are obtained for greater post-selected signal photo-count numbers, because of the low number of measurements that were performed on individual photo-count numbers of fields with broader photo-count distributions. In other words, measurement imperfections cause local variations of photo-count distributions that give artificially larger values of NCDs, especially in the area with higher photo-count numbers, $c_{\rm i}$ [see~Figs.~3(b) and~5(b) in Ref.~\cite{PerinaJr2020}]. Notably, both these measurement effects are removed when the NCDs are determined by our designed ANN. We remark that elimination of these effects can also be accomplished by using the FIT method. However, to perform the proper fitting, data belonging to all post-selected idler fields have to be processes together. This task is not needed once our ANN is trained.

\begin{figure*}[t!]
\centering
	\includegraphics{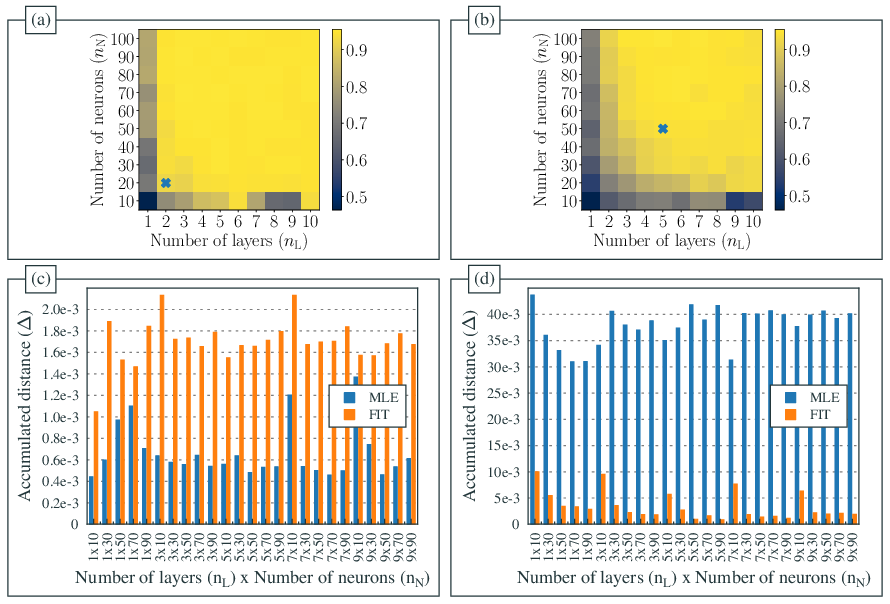}
	\caption{(a,b) Artificial neural network mean accuracy and (c,d) accumulated distance~$\Delta = \sum [\tau_{i}^{\rm (A)}-\tau_{i}^{(R)}]^{2}$ between ANN predictions~[$\tau^{(A)}$] and NCDs estimated using maximal likelihood estimation~(MLE) and Gaussian twin-beam fit~(FIT) reconstruction techniques [$\tau^{(R)}$]. (a) and (c) show the results for intensity-moment evaluations, while (b) and (d) stand for probability calculations. The X- and Y-axes [in~(a) and~(b)] and X-axis [in~(c) and~(d)] of the plots correspond to ANN's number of layers~($n_{\rm L}$) and neurons~($n_{\rm N}$). The blue $\times$-marks in figures~(a) and~(b) show the network designs used in our final nonclassicality classifier.}
\label{Figure6}
\end{figure*}

We conclude this section by pointing out that several other network designs and hyper-parameter configurations were tested; however, they showed no significant benefit compared to the final versions used in this study. Indeed, by increasing the number of hidden layers~($n_{L}$) and/or neurons~($n_{N}$) per layer over certain value, the networks had tendency to over-parameterize the problem, which resulted in poor efficiency when processing the experimental data. The~test accuracy (i.e.~the relative number of succesful classifications) determined after 250~training epochs is shown in Fig.~\ref{Figure6}(a, b). It identifies the general parameters of successful ANNs. The bar-plots in Fig.~\ref{Figure6}(c, d) reveals another important property of the tested ANNs: when compared with the experimental results, represented by MLE reconstruction and the FIT method, the ANN's results are always closer to those coming from the MLE reconstruction for moment based NCWs, while the opposite is true for probability based NCWs. This reflects the fact that ANNs are able to predict the NCDs for individual post-selected idler fields, i.e. without taking into account the properties of neighbor post-selected fields (as it is the case of the Gaussian FIT model); therefore, they can reveal the locality of the nonclassicality. More importantly, they are not overpowered by artificial increase of local nonclassicality caused by inaccurate estimation of particular photo-count histogram values due to experimental imperfections.

\section{Discussion} 

Artificial neural networks have been found useful in the quantification of nonclassicality of potentially sub-Poissonian beams. They allow to bypass the reconstruction of beam photon-number distributions using, e.g.~the MLE reconstruction method. This results in fast assignment of characteristic nonclassicality depths obtained by applying suitable intensity-moment, and probability based nonclassicality witnesses. Artificial neural networks allow to automatically involve broad groups of nonclassicality witnesses in the process of their learning, thus providing an~efficient performance in determining nonclassicality metrics, such as the nonclassicality depth. It has been demonstrated that the networks work with the measurement~(detection) noises in qualitatively better way compared to the MLE reconstruction. More specifically, ANNs are able to judge nonclassicality regardless of the local influence of noise, that either affects the measurement of individual photo-count probabilities or modifies the photo-count probabilities in small areas. When using the MLE algorithm, the local noise may result either in the artificial increase or decrease of the nonclassicality depth. On the other hand, the Gaussian best-fit procedure is able to cope with this local noise, however, only at the expense of combining different measurements together. Thus, artificial neural networks give us qualitatively better tools for estimating the beam nonclassicality compared to the commonly-used methods.

\section{Methods} 

\noindent \textbf{Mapping between intensity moments and photon-number probabilities}

\noindent The Mandel detection formula gives the probability~$p_{\rm i}(n)$ of detecting~$n$ photons when a~detector is illuminated by the field described by the quasi-distribution~$P_{\rm i}(W)$ of integrated intensity~$W$ related to the normal ordering of field operators~\cite{Perina1991,Mandel1995}:
\begin{equation}
	p(n) = \frac{1}{n!} \int_{0}^{\infty} dW \, W^{n} \exp(-W) P_{\rm i}(W).
\label{eq9}
\end{equation}
This formula can be rewritten as follows
\begin{equation}
	\frac{n! p_{\rm i}(n)}{p_{\rm i}(0)} =
	\frac{\int_{0}^{\infty} dW \, W^{n} \exp(-W) P_{\rm i}(W)}
		{\int_{0}^{\infty} dW \, \exp(-W) P_{\rm i}(W) } \equiv
	\langle W^{n} \rangle_{\rm mod}
\label{eq10}
\end{equation}
in which the right-hand side gives intensity moments~$\langle W^{n} \rangle_{\rm mod}$ of a~properly-normalized quasi-distribution. Once we write the intensity NCWs using the modified moments~$\langle W^{n} \rangle_{\rm mod}$, we may map these intensity NCWs into the probability NCWs using equation~(\ref{eq10})~\cite{Klyshko1996,Waks2004,Waks2006,Wakui2014,PerinaJr2017a}:
\begin{equation}
	\langle W^{n} \rangle_{\rm mod} \longleftarrow
		\frac{n! p_{\rm i}(n)}{p_{\rm i}(0)}.
\label{eq11}
\end{equation}
This mapping immediately transforms the intensity NCWs given in equation~(\ref{eq3}) into the probability NCWs written in equation~(\ref{eq4}).
\\

\noindent \textbf{Transformation of intensity moments and photon-number distributions into an arbitrary field-operator ordering}

\noindent By realizing that intensity moments~$\langle W^{k'} \rangle$, with $k'=1,\ldots$ are related to the normal ordering of field operators, one can determine the intensity moments~$\langle W^{k} \rangle_{s}$, $ k=1,\ldots$ for a~general field-operator ordering parameter~$s$~(with~$s \in \langle-1,1\rangle$), and a~field comprising~$M$ independent modes, by making use of the expression~\cite{Perina1991,PerinaJr2022}:
\begin{equation}
	\langle W^{k} \rangle_{s} = \sum_{k'=1}^{k} S_W(k,k';s,M)
		\langle W^{k'} \rangle,
\label{eq12}
\end{equation}
where
\begin{equation}
	S_{W}(k,k';s,M) = \binom{k}{k'} \frac{\Gamma(k+M)}{\Gamma(k'+M)}
		\left( \frac{1-s}{2} \right)^{k-k'}.
\label{eq13}
\end{equation}
Similarly, the usual photon-number distribution~$p_{\rm i}$ (in the normal field-operator ordering) is transformed into its~$s$-ordered counterpart~$p_{{\rm i},s}$ with the help of the following relation~\cite{PerinaJr2020a,PerinaJr2022}:
\begin{equation}
	p_{{\rm i},s}(n) = \sum_{n'=0}^{\infty} S_{p}(n,n';s,M) p_{\rm i}(n'),
\label{eq14}
\end{equation}
with
\begin{eqnarray}
	S_{p}(n,n';s,M) &=& \left( \frac{2}{3-s} \right)^M
		\left( \frac{1+s}{1-s} \right)^{n'}
		\left( \frac{1-s}{3-s} \right)^{n} \nonumber \\
	& \times & \sum_{l=0}^{n'} (-1)^{n'-l} \binom{n'}{l}
		\binom{n+l+M-1}{n} \nonumber \\
	& \times & \left( \frac{4}{(1+s)(3-s)} \right)^{l}.
\label{eq15}
\end{eqnarray}

Finally, we note that the intensity moments~$\langle W^{k} \rangle$, that are the normally-ordered photon-number moments, are derived from the usual photon-number moments~$\langle n^{k'} \rangle$ using the Stirling numbers~$S$ of the first kind~\cite{Gradshtein2000}:
\begin{equation}
	\langle W^{k} \rangle = \sum_{k'=0}^{k} S(k,k') \langle n^{k'} \rangle.
\label{eq16}
\end{equation}


\vspace{3mm} \noindent \textbf{Data Availability.}
Data underlying the results presented in this work may be obtained from the authors upon reasonable request.

\vspace{3mm} \noindent \textbf{Acknowledgements.}
R.M, J.P., V.M., and O.H. acknowledge support by the project OP JAC CZ.02.01.01/00/22\_008/0004596 of the Ministry of Education, Youth, and Sports of the Czech Republic and EU. R.J.L.-M. thankfully acknowledges financial support by DGAPA-UNAM under the project UNAM-PAPIIT IN101623.

\vspace{3mm} \noindent \textbf{Conflict of Interest.}
The authors declare no conflicts of interest regarding this article.

\vspace{3mm} \noindent \textbf{Author Contributions.}
R.J.L.-M., R.M., and J.P. conceived the experiment, O.H., R.M., and V.M. carried out the experiment, R.J.L.-M. and R.M. (J.P.) processed the experimental data from the point of view of AI (nonclassicality). All authors contributed to the interpretation of the results and to the preparation of the manuscript.


\bibliographystyle{naturemag}
\bibliography{machulka}

\begin{thebibliography}{10}
\expandafter\ifx\csname url\endcsname\relax
  \def\url#1{\texttt{#1}}\fi
\expandafter\ifx\csname urlprefix\endcsname\relax\def\urlprefix{URL }\fi
\providecommand{\bibinfo}[2]{#2}
\providecommand{\eprint}[2][]{\url{#2}}

\bibitem{Glauber1963}
\bibinfo{author}{Glauber, R.~J.}
\newblock \bibinfo{title}{Coherent and incoherent states of the radiation
  field}.
\newblock \emph{\bibinfo{journal}{Phys. Rev.}} \textbf{\bibinfo{volume}{131}},
  \bibinfo{pages}{2766---2788} (\bibinfo{year}{1963}).

\bibitem{Sudarshan1963}
\bibinfo{author}{Sudarshan, E. C.~G.}
\newblock \bibinfo{title}{Equivalence of semiclassical and quantum mechanical
  descriptions of statistical light beams}.
\newblock \emph{\bibinfo{journal}{Phys. Rev. Lett.}}
  \textbf{\bibinfo{volume}{10}}, \bibinfo{pages}{277---179}
  (\bibinfo{year}{1963}).

\bibitem{Mandel1995}
\bibinfo{author}{Mandel, L.} \& \bibinfo{author}{Wolf, E.}
\newblock \emph{\bibinfo{title}{Optical Coherence and Quantum Optics}}
  (\bibinfo{publisher}{Cambridge University, Cambridge}, \bibinfo{year}{1995}).

\bibitem{Dodonov2002}
\bibinfo{author}{Dodonov, V.~V.}
\newblock \bibinfo{title}{Nonclassical states in quantum optics: A squeezed
  review of the first 75 years}.
\newblock \emph{\bibinfo{journal}{J. Opt. B: Quantum Semiclass. Opt.}}
  \textbf{\bibinfo{volume}{4}}, \bibinfo{pages}{R1---R33}
  (\bibinfo{year}{2002}).

\bibitem{Luks1988}
\bibinfo{author}{Luk\v{s}, A.}, \bibinfo{author}{Pe\v{r}inov\'{a}, V.} \&
  \bibinfo{author}{Pe\v{r}ina, J.}
\newblock \bibinfo{title}{Principal squeezing of vacuum fluctuations}.
\newblock \emph{\bibinfo{journal}{Opt. Commun.}} \textbf{\bibinfo{volume}{67}},
  \bibinfo{pages}{149---151} (\bibinfo{year}{1988}).

\bibitem{Hill1997}
\bibinfo{author}{Hill, S.} \& \bibinfo{author}{Wootters, W.~K.}
\newblock \bibinfo{title}{Computable entanglement}.
\newblock \emph{\bibinfo{journal}{Phys. Rev. Lett.}}
  \textbf{\bibinfo{volume}{78}}, \bibinfo{pages}{5022---5025}
  (\bibinfo{year}{1997}).

\bibitem{Adesso2007}
\bibinfo{author}{Adesso, G.} \& \bibinfo{author}{Illuminati, F.}
\newblock \bibinfo{title}{Entanglement in continuous variable systems: Recent
  advances and current perspectives}.
\newblock \emph{\bibinfo{journal}{J. Phys. A: Math. Theor.}}
  \textbf{\bibinfo{volume}{40}}, \bibinfo{pages}{7821---7880}
  (\bibinfo{year}{2007}).

\bibitem{Lvovsky2009}
\bibinfo{author}{Lvovsky, A.~I.} \& \bibinfo{author}{Raymer, M.~G.}
\newblock \bibinfo{title}{Continuous-variable optical quantum state
  tomography}.
\newblock \emph{\bibinfo{journal}{Rev. Mod. Phys.}}
  \textbf{\bibinfo{volume}{81}}, \bibinfo{pages}{299---332}
  (\bibinfo{year}{2009}).

\bibitem{Li2020}
\bibinfo{author}{Li, T.}, \bibinfo{author}{Li, F.}, \bibinfo{author}{Altuzarra,
  C.}, \bibinfo{author}{Classen, A.} \& \bibinfo{author}{Agarwal, G.~S.}
\newblock \bibinfo{title}{{Squeezed light induced two-photon absorption
  fluorescence of fluorescein biomarkers}}.
\newblock \emph{\bibinfo{journal}{Applied Physics Letters}}
  \textbf{\bibinfo{volume}{116}}, \bibinfo{pages}{254001}
  (\bibinfo{year}{2020}).

\bibitem{Genovese2016}
\bibinfo{author}{Genovese, M.}
\newblock \bibinfo{title}{Real applications of quantum imaging}.
\newblock \emph{\bibinfo{journal}{J. Opt.}} \textbf{\bibinfo{volume}{18}},
  \bibinfo{pages}{073002} (\bibinfo{year}{2016}).

\bibitem{Zeshen2023}
\bibinfo{author}{Zhang, Z.} \emph{et~al.}
\newblock \bibinfo{title}{Entanglement-based quantum information technology}
  (\bibinfo{year}{2023}).
\newblock \eprint{2308.01416}.

\bibitem{Klyshko1980}
\bibinfo{author}{Klyshko, D.~N.}
\newblock \bibinfo{title}{Use of two-photon light for absolute calibration of
  photoelectric detectors}.
\newblock \emph{\bibinfo{journal}{Sov. J. Quantum Electron.}}
  \textbf{\bibinfo{volume}{10}}, \bibinfo{pages}{1112} (\bibinfo{year}{1980}).

\bibitem{Perina1990}
\bibinfo{author}{Pe\v{r}ina, J.} \& \bibinfo{author}{Bajer, J.}
\newblock \bibinfo{title}{Origin of oscillations in photon distributions of
  squeezed states}.
\newblock \emph{\bibinfo{journal}{Phys. Rev. A}} \textbf{\bibinfo{volume}{41}},
  \bibinfo{pages}{516---518} (\bibinfo{year}{1990}).

\bibitem{Kenfack2004}
\bibinfo{author}{Kenfack, A.} \& \bibinfo{author}{Zyczkowski, K.}
\newblock \bibinfo{title}{Negativity of the {Wigner} function as an indicator
  of nonclassicality}.
\newblock \emph{\bibinfo{journal}{J. Opt. B: Quantum Semiclass. Opt.}}
  \textbf{\bibinfo{volume}{6}}, \bibinfo{pages}{396---404}
  (\bibinfo{year}{2004}).

\bibitem{Richter2002}
\bibinfo{author}{Richter, T.} \& \bibinfo{author}{Vogel, W.}
\newblock \bibinfo{title}{Nonclassicality of quantum states: A hierarchy of
  observable conditions}.
\newblock \emph{\bibinfo{journal}{Phys. Rev. Lett.}}
  \textbf{\bibinfo{volume}{89}}, \bibinfo{pages}{283601}
  (\bibinfo{year}{2002}).

\bibitem{Miranowicz2010}
\bibinfo{author}{Miranowicz, A.}, \bibinfo{author}{Bartkowiak, M.},
  \bibinfo{author}{Wang, X.}, \bibinfo{author}{Liu, Y.-X.} \&
  \bibinfo{author}{Nori, F.}
\newblock \bibinfo{title}{Testing nonclassicality in multimode fields: A
  unified derivation of classical inequalities}.
\newblock \emph{\bibinfo{journal}{Phys. Rev. A}} \textbf{\bibinfo{volume}{82}},
  \bibinfo{pages}{013824} (\bibinfo{year}{2010}).

\bibitem{Ryl2015}
\bibinfo{author}{Ryl, S.} \emph{et~al.}
\newblock \bibinfo{title}{Unified nonclassicality criteria}.
\newblock \emph{\bibinfo{journal}{Phys. Rev. A}} \textbf{\bibinfo{volume}{92}},
  \bibinfo{pages}{011801(R)} (\bibinfo{year}{2015}).

\bibitem{agarwal1988}
\bibinfo{author}{Agarwal, G.~S.}
\newblock \bibinfo{title}{Nonclassical statistics of fields in pair coherent
  states}.
\newblock \emph{\bibinfo{journal}{J. Opt. Soc. Am. B}}
  \textbf{\bibinfo{volume}{5}}, \bibinfo{pages}{1940--1947}
  (\bibinfo{year}{1988}).

\bibitem{PerinaJr2017a}
\bibinfo{author}{{Pe\v{r}ina~Jr.}, J.}, \bibinfo{author}{Arkhipov, I.~I.},
  \bibinfo{author}{{Mich\' alek}, V.} \& \bibinfo{author}{Haderka, O.}
\newblock \bibinfo{title}{Non-classicality and entanglement criteria for
  bipartite optical fields characterized by quadratic detectors}.
\newblock \emph{\bibinfo{journal}{Phys. Rev. A}} \textbf{\bibinfo{volume}{96}},
  \bibinfo{pages}{043845} (\bibinfo{year}{2017}).

\bibitem{PerinaJr2020a}
\bibinfo{author}{{Pe\v{r}ina~Jr.}, J.}, \bibinfo{author}{Haderka, O.} \&
  \bibinfo{author}{{Mich\'{a}lek}, V.}
\newblock \bibinfo{title}{Non-classicality and entanglement criteria for
  bipartite optical fields characterized by quadratic detectors {II}: Criteria
  based on probabilities}.
\newblock \emph{\bibinfo{journal}{Phys. Rev. A}}
  \textbf{\bibinfo{volume}{102}}, \bibinfo{pages}{043713}
  (\bibinfo{year}{2020}).

\bibitem{PerinaJr2022}
\bibinfo{author}{{Pe\v{r}ina~Jr.}, J.}, \bibinfo{author}{{Pavl\' \i\v{c}ek},
  P.}, \bibinfo{author}{{Mich\'{a}lek}, V.}, \bibinfo{author}{Machulka, R.} \&
  \bibinfo{author}{Haderka, O.}
\newblock \bibinfo{title}{Nonclassicality criteria for n-dimensional optical
  fields detected by quadratic detectors}.
\newblock \emph{\bibinfo{journal}{Phys. Rev. A}}
  \textbf{\bibinfo{volume}{105}}, \bibinfo{pages}{013706}
  (\bibinfo{year}{2022}).

\bibitem{Perina1991}
\bibinfo{author}{Pe\v{r}ina, J.}
\newblock \emph{\bibinfo{title}{Quantum Statistics of Linear and Nonlinear
  Optical Phenomena}} (\bibinfo{publisher}{Kluwer, Dordrecht},
  \bibinfo{year}{1991}).

\bibitem{Klyshko1996}
\bibinfo{author}{Klyshko, D.~N.}
\newblock \bibinfo{title}{Observable signs of nonclassical light}.
\newblock \emph{\bibinfo{journal}{Phys. Lett. A}}
  \textbf{\bibinfo{volume}{213}}, \bibinfo{pages}{7---15}
  (\bibinfo{year}{1996}).

\bibitem{Lee1991}
\bibinfo{author}{Lee, C.~T.}
\newblock \bibinfo{title}{Measure of the nonclassicality of nonclassical
  states}.
\newblock \emph{\bibinfo{journal}{Phys. Rev. A}} \textbf{\bibinfo{volume}{44}},
  \bibinfo{pages}{R2775---R2778} (\bibinfo{year}{1991}).

\bibitem{PerinaJr2019}
\bibinfo{author}{{Pe\v{r}ina~Jr.}, J.}, \bibinfo{author}{Haderka, O.} \&
  \bibinfo{author}{{Mich\' alek}, V.}
\newblock \bibinfo{title}{Simultaneous observation of higher-order
  non-classicalities based on experimental photocount moments and
  probabilities}.
\newblock \emph{\bibinfo{journal}{Sci. Rep.}} \textbf{\bibinfo{volume}{9}},
  \bibinfo{pages}{8961} (\bibinfo{year}{2019}).

\bibitem{PerinaJr2021}
\bibinfo{author}{{Pe\v{r}ina~Jr.}, J.}, \bibinfo{author}{{Mich\'{a}lek}, V.},
  \bibinfo{author}{Machulka, R.} \& \bibinfo{author}{Haderka, O.}
\newblock \bibinfo{title}{Two-beam light with simultaneous anti-correlations in
  photon-number fluctuations and {sub-Poissonian} statistics}.
\newblock \emph{\bibinfo{journal}{Phys. Rev. A}}
  \textbf{\bibinfo{volume}{104}}, \bibinfo{pages}{013712}
  (\bibinfo{year}{2021}).

\bibitem{PerinaJr2021a}
\bibinfo{author}{{Pe\v{r}ina~Jr.}, J.}, \bibinfo{author}{{Mich\'{a}lek}, V.},
  \bibinfo{author}{Machulka, R.} \& \bibinfo{author}{Haderka, O.}
\newblock \bibinfo{title}{Two-beam light with 'checkered-pattern' photon-number
  distributions}.
\newblock \emph{\bibinfo{journal}{Opt. Express}} \textbf{\bibinfo{volume}{29}},
  \bibinfo{pages}{29704} (\bibinfo{year}{2021}).

\bibitem{ahmed2021}
\bibinfo{author}{Ahmed, S.}, \bibinfo{author}{S\'anchez Mu\~noz, C.},
  \bibinfo{author}{Nori, F.} \& \bibinfo{author}{Kockum, A.~F.}
\newblock \bibinfo{title}{Classification and reconstruction of optical quantum
  states with deep neural networks}.
\newblock \emph{\bibinfo{journal}{Phys. Rev. Res.}}
  \textbf{\bibinfo{volume}{3}}, \bibinfo{pages}{033278} (\bibinfo{year}{2021}).

\bibitem{you1}
\bibinfo{author}{You, C.} \emph{et~al.}
\newblock \bibinfo{title}{{Identification of light sources using machine
  learning}}.
\newblock \emph{\bibinfo{journal}{Applied Physics Reviews}}
  \textbf{\bibinfo{volume}{7}}, \bibinfo{pages}{021404} (\bibinfo{year}{2020}).

\bibitem{gebhart2021}
\bibinfo{author}{Gebhart, V.} \emph{et~al.}
\newblock \bibinfo{title}{Identifying nonclassicality from experimental data
  using artificial neural networks}.
\newblock \emph{\bibinfo{journal}{Phys. Rev. Res.}}
  \textbf{\bibinfo{volume}{3}}, \bibinfo{pages}{023229} (\bibinfo{year}{2021}).

\bibitem{Bhusal2022}
\bibinfo{author}{Bhusal, N.} \emph{et~al.}
\newblock \bibinfo{title}{Smart quantum statistical imaging beyond the
  abbe-rayleigh criterion}.
\newblock \emph{\bibinfo{journal}{npj Quantum Information}}
  \textbf{\bibinfo{volume}{8}}, \bibinfo{pages}{83} (\bibinfo{year}{2022}).

\bibitem{hsieh2022}
\bibinfo{author}{Hsieh, H.-Y.} \emph{et~al.}
\newblock \bibinfo{title}{Extract the degradation information in squeezed
  states with machine learning}.
\newblock \emph{\bibinfo{journal}{Phys. Rev. Lett.}}
  \textbf{\bibinfo{volume}{128}}, \bibinfo{pages}{073604}
  (\bibinfo{year}{2022}).

\bibitem{gray2018}
\bibinfo{author}{Gray, J.}, \bibinfo{author}{Banchi, L.},
  \bibinfo{author}{Bayat, A.} \& \bibinfo{author}{Bose, S.}
\newblock \bibinfo{title}{Machine-learning-assisted many-body entanglement
  measurement}.
\newblock \emph{\bibinfo{journal}{Phys. Rev. Lett.}}
  \textbf{\bibinfo{volume}{121}}, \bibinfo{pages}{150503}
  (\bibinfo{year}{2018}).
\newblock
  \urlprefix\url{https://link.aps.org/doi/10.1103/PhysRevLett.121.150503}.

\bibitem{Asif2023}
\bibinfo{author}{Asif, N.}, \bibinfo{author}{Khalid, U.},
  \bibinfo{author}{Khan, A.}, \bibinfo{author}{Duong, T.~Q.} \&
  \bibinfo{author}{Shin, H.}
\newblock \bibinfo{title}{Entanglement detection with artificial neural
  networks}.
\newblock \emph{\bibinfo{journal}{Scientific Reports}}
  \textbf{\bibinfo{volume}{13}}, \bibinfo{pages}{1562} (\bibinfo{year}{2023}).

\bibitem{koutny2023}
\bibinfo{author}{Koutný, D.} \emph{et~al.}
\newblock \bibinfo{title}{Deep learning of quantum entanglement from incomplete
  measurements}.
\newblock \emph{\bibinfo{journal}{Science Advances}}
  \textbf{\bibinfo{volume}{9}}, \bibinfo{pages}{eadd7131}
  (\bibinfo{year}{2023}).

\bibitem{PerinaJr2013b}
\bibinfo{author}{{Pe\v{r}ina~Jr.}, J.}, \bibinfo{author}{Haderka, O.} \&
  \bibinfo{author}{Mich\'{a}lek, V.}
\newblock \bibinfo{title}{{Sub-Poissonian-light generation} by postselection
  from twin beams}.
\newblock \emph{\bibinfo{journal}{Opt. Express}} \textbf{\bibinfo{volume}{21}},
  \bibinfo{pages}{19387---19394} (\bibinfo{year}{2013}).

\bibitem{Arkhipov2016c}
\bibinfo{author}{Arkhipov, I.~I.}, \bibinfo{author}{{Pe\v{r}ina~Jr.}, J.},
  \bibinfo{author}{Mich\'{a}lek, V.} \& \bibinfo{author}{Haderka, O.}
\newblock \bibinfo{title}{Experimental detection of nonclassicality of
  single-mode fields via intensity moments}.
\newblock \emph{\bibinfo{journal}{Opt. Express}} \textbf{\bibinfo{volume}{24}},
  \bibinfo{pages}{29496---29505} (\bibinfo{year}{2016}).

\bibitem{PerinaJr2017c}
\bibinfo{author}{{Pe\v{r}ina~Jr.}, J.}, \bibinfo{author}{{Mich\' alek}, V.} \&
  \bibinfo{author}{Haderka, O.}
\newblock \bibinfo{title}{Higher-order {sub-Poissonian-like} nonclassical
  fields: Theoretical and experimental comparison}.
\newblock \emph{\bibinfo{journal}{Phys. Rev. A}} \textbf{\bibinfo{volume}{96}},
  \bibinfo{pages}{033852} (\bibinfo{year}{2017}).

\bibitem{Laurat2003}
\bibinfo{author}{Laurat, J.}, \bibinfo{author}{Coudreau, T.},
  \bibinfo{author}{Treps, N.}, \bibinfo{author}{Maitre, A.} \&
  \bibinfo{author}{Fabre, C.}
\newblock \bibinfo{title}{Conditional preparation of a quantum state in the
  continuous variable regime: Generation of a {sub-Poissonian} state from twin
  beams}.
\newblock \emph{\bibinfo{journal}{Phys. Rev. Lett.}}
  \textbf{\bibinfo{volume}{91}}, \bibinfo{pages}{213601}
  (\bibinfo{year}{2003}).

\bibitem{Brida2012}
\bibinfo{author}{Brida, G.} \emph{et~al.}
\newblock \bibinfo{title}{An extremely low-noise heralded single-photon source:
  A breakthrough for quantum technologies}.
\newblock \emph{\bibinfo{journal}{Appl. Phys. Lett.}}
  \textbf{\bibinfo{volume}{101}} (\bibinfo{year}{2012}).

\bibitem{Haderka2005}
\bibinfo{author}{Haderka, O.}, \bibinfo{author}{{Pe\v{r}ina~Jr.}, J.} \&
  \bibinfo{author}{Hamar, M.}
\newblock \bibinfo{title}{Simple direct measurement of nonclassical joint
  signal-idler photon-number statistics and correlation area of twin photon
  beams}.
\newblock \emph{\bibinfo{journal}{J. Opt. B: Quantum Semiclass. Opt.}}
  \textbf{\bibinfo{volume}{7}}, \bibinfo{pages}{S572---S576}
  (\bibinfo{year}{2005}).

\bibitem{PerinaJr2012}
\bibinfo{author}{{Pe\v{r}ina~Jr.}, J.}, \bibinfo{author}{Hamar, M.},
  \bibinfo{author}{Mich\'{a}lek, V.} \& \bibinfo{author}{Haderka, O.}
\newblock \bibinfo{title}{Photon-number distributions of twin beams generated
  in spontaneous parametric down-conversion and measured by an intensified
  {CCD} camera}.
\newblock \emph{\bibinfo{journal}{Phys. Rev. A}} \textbf{\bibinfo{volume}{85}},
  \bibinfo{pages}{023816} (\bibinfo{year}{2012}).

\bibitem{Dempster1977}
\bibinfo{author}{Dempster, A.~P.}, \bibinfo{author}{Laird, N.~M.} \&
  \bibinfo{author}{Rubin, D.~B.}
\newblock \bibinfo{title}{Maximum likelihood from incomplete data via the {EM}
  algorithm}.
\newblock \emph{\bibinfo{journal}{J. Royal Statist. Soc. B}}
  \textbf{\bibinfo{volume}{39}}, \bibinfo{pages}{1---38}
  (\bibinfo{year}{1977}).

\bibitem{PerinaJr2020}
\bibinfo{author}{{Pe\v{r}ina~Jr.}, J.}, \bibinfo{author}{{Mich\'{a}lek}, V.} \&
  \bibinfo{author}{Haderka, O.}
\newblock \bibinfo{title}{Non-classicality of optical fields as observed in
  photocount and photon-number distributions}.
\newblock \emph{\bibinfo{journal}{Opt. Express}} \textbf{\bibinfo{volume}{28}},
  \bibinfo{pages}{32620--32631} (\bibinfo{year}{2020}).

\bibitem{PerinaJr2013a}
\bibinfo{author}{{Pe\v{r}ina~Jr.}, J.}, \bibinfo{author}{Haderka, O.},
  \bibinfo{author}{Mich\'{a}lek, V.} \& \bibinfo{author}{Hamar, M.}
\newblock \bibinfo{title}{State reconstruction of a multimode twin beam using
  photodetection}.
\newblock \emph{\bibinfo{journal}{Phys. Rev. A}} \textbf{\bibinfo{volume}{87}},
  \bibinfo{pages}{022108} (\bibinfo{year}{2013}).

\bibitem{PerinaJr2012a}
\bibinfo{author}{{Pe\v{r}ina~Jr.}, J.}, \bibinfo{author}{Haderka, O.},
  \bibinfo{author}{Hamar, M.} \& \bibinfo{author}{Mich\'{a}lek, V.}
\newblock \bibinfo{title}{Absolute detector calibration using twin beams}.
\newblock \emph{\bibinfo{journal}{Opt. Lett.}} \textbf{\bibinfo{volume}{37}},
  \bibinfo{pages}{2475---2477} (\bibinfo{year}{2012}).

\bibitem{Waks2004}
\bibinfo{author}{Waks, E.}, \bibinfo{author}{Diamanti, E.},
  \bibinfo{author}{Sanders, B.~C.}, \bibinfo{author}{Bartlett, S.~D.} \&
  \bibinfo{author}{Yamamoto, Y.}
\newblock \bibinfo{title}{Direct observation of nonclassical photon statistics
  in parametric down-conversion}.
\newblock \emph{\bibinfo{journal}{Phys. Rev. Lett.}}
  \textbf{\bibinfo{volume}{92}}, \bibinfo{pages}{113602}
  (\bibinfo{year}{2004}).

\bibitem{Waks2006}
\bibinfo{author}{Waks, E.}, \bibinfo{author}{Sanders, B.~C.},
  \bibinfo{author}{Diamanti, E.} \& \bibinfo{author}{Yamamoto, Y.}
\newblock \bibinfo{title}{Highly nonclassical photon statistics in parametric
  down-conversion}.
\newblock \emph{\bibinfo{journal}{Phys. Rev. A}} \textbf{\bibinfo{volume}{73}},
  \bibinfo{pages}{033814} (\bibinfo{year}{2006}).

\bibitem{Wakui2014}
\bibinfo{author}{Wakui, K.} \emph{et~al.}
\newblock \bibinfo{title}{Ultrabroadband direct detection of nonclassical
  photon statistics at telecom wavelength}.
\newblock \emph{\bibinfo{journal}{Sci. Rep.}} \textbf{\bibinfo{volume}{4}},
  \bibinfo{pages}{4535} (\bibinfo{year}{2014}).

\bibitem{Gradshtein2000}
\bibinfo{author}{Gradshtein, I.~S.} \& \bibinfo{author}{Ryzhik, I.~M.}
\newblock \emph{\bibinfo{title}{Table of Integrals, Series, and Products, 6th
  ed.}} (\bibinfo{publisher}{Academic Press, San Diego}, \bibinfo{year}{2000}).

\end{thebibliography}

\end{document}